\documentclass[a4paper,fleqn,usenatbib, letters, useAMS]{mnras}
\usepackage{graphicx}
\usepackage{multicol}
\usepackage{enumerate}
\usepackage{amsmath}
\usepackage{pdflscape}
\usepackage{amssymb}
\usepackage{longtable}

\usepackage{journal_names}
\def\farcs{\hbox{$.\!\!^{\prime\prime}$}}

\title[CX361: a new high state AM CVn] {Discovery of a high state AM CVn binary in the Galactic Bulge Survey}
\author[Wevers et al.]{T.~Wevers$^{1}$\thanks{Email: t.wevers@astro.ru.nl}\thanks{Based on observations collected at the European Organisation for Astronomical Research in the Southern Hemisphere under ESO programme 091.D-0062(A).}, M.\,A.\,P.~Torres$^{2,1}$, P.\,G.~Jonker$^{2,1}$, J.\,D.~Wetuski$^{3}$, G.~Nelemans$^{1,4}$,
\newauthor D. Steeghs$^{5}$, T.\,J.~Maccarone$^{6}$, C.~Heinke$^{7}$, R.\,I.~Hynes$^{3}$, A.~Udalski$^{8}$, \newauthor Z.~Kostrzewa-Rutkowska$^{2,1}$, P.\,J.~Groot$^{1}$, R.~Gazer$^{7}$, M.\,K.~Szyma{\'n}ski$^{8}$, C. T. Britt$^{6}$, \newauthor \L{}.~Wyrzykowski$^{8}$, R. Poleski$^{9}$  \\\\
$^{1}$Department of Astrophysics/IMAPP, Radboud University, P.O. Box 9010, 6500 GL Nijmegen, the Netherlands\\
$^{2}$SRON, Netherlands Institute for Space Research, Sorbonnelaan 2, 3584 CA Utrecht, the Netherlands\\
$^{3}$Department of Physics and Astronomy, Louisiana State University, Baton Rouge, LA 70803-4001, USA\\
$^{4}$Institute for Astronomy, KU Leuven, Celestijnenlaan 200D, 3001 Leuven, Belgium\\
$^{5}$Department of Physics, University of Warwick, Coventry CV4 7AL, UK\\
$^{6}$Department of Physics, Texas Tech University, box 41051, Lubbock, TX 79409-1051, USA\\
$^{7}$Dept. of Physics, University of Alberta, CCLS 4-183, Edmonton, AB T6G 2E1, Canada\\
$^{8}$Warsaw University Astronomical Observatory, Al. Ujazdowskie 4, PL-00-478 Warszawa, Poland\\
$^{9}$Department of Astronomy, Ohio State University, 140 W. 18th Ave., Columbus, OH 43210, USA\\
}

\begin{document}
\date{Accepted 2016 July 13. Received 2016 July 12; in original form 2016 June 24}
\pagerange{\pageref{firstpage}--\pageref{lastpage}} \pubyear{2016}
\maketitle
\label{firstpage}

\begin{abstract}
We report on the discovery of a hydrogen-deficient compact binary (CXOGBS J175107.6-294037) belonging to the AM CVn class in the Galactic Bulge Survey. Deep archival X-ray observations constrain the X-ray positional uncertainty of the source to 0\farcs57, and allow us to uniquely identify the optical and UV counterpart. Optical spectroscopic observations reveal the presence of broad, shallow He \textsc{i} absorption lines while no sign of hydrogen is present, consistent with a high state system. We present the optical lightcurve from Optical Gravitational Lensing Experiment monitoring, spanning 15 years. It shows no evidence for outbursts; variability is present at the 0.2 mag level on timescales ranging from hours to weeks. A modulation on a timescale of years is also observed. A Lomb-Scargle analysis of the optical lightcurves shows two significant periodicities at 22.90 and 23.22 min. Although the physical interpretation is uncertain, such timescales are in line with expectations for the orbital and superhump periods. We estimate the distance to the source to be between 0.5\,--\,1.1 kpc. Spectroscopic follow-up observations are required to establish the orbital period, and to determine whether this source can serve as a verification binary for the \textit{eLISA} gravitational wave mission.
\end{abstract}

\begin{keywords}
X-rays: binaries -- stars: individual: CXOGBS J175107.6-294037 -- white dwarfs -- binaries: close -- stars: evolution -- accretion discs
\end{keywords}


\section{Introduction}
\label{sec:introduction}
AM Canum Venaticorum (AM CVn) systems are a class of binary stars, consisting of a white dwarf (WD) accreting H-deficient material from a low mass companion. These binaries are in a very compact configuration with observed orbital periods ranging from 5 to 65 min, suggesting the companion is a (semi-) degenerate star (see e.g. \citeauthor{Solheim2010} \citeyear{Solheim2010} for a recent review). 
The class of AM CVn stars can be divided into four groups, depending on the evolutionary stage of the binary:
\begin{enumerate}
\item P$_{\text{orb}} \lesssim$ 10 min: direct impact systems without a disc
\item 10 min $\lesssim$ P$_{\text{orb}} \lesssim$ 20 min: stable disc in a persistent high state
\item 20 min $\lesssim$ P$_{\text{orb}} \lesssim$ 40 min: variable disc with outbursts
\item 40 min $\lesssim$ P$_{\text{orb}} \lesssim$ 65 min: stable disc in quiescence
 \end{enumerate} 
 The orbital evolution of these short period systems is expected to be dominated by gravitational wave radiation (GWR; \citeauthor{Tutukov1979} \citeyear{Tutukov1979}). In that case, the mass transfer rate is set by angular momentum loss due to GWR, and because the latter is a steep function of the orbital period, this implies a strong decrease of $\dot{M}$ with increasing P$_{orb}$. This provides a natural explanation for the observed behaviour. With decreasing $\dot{M}$, the evolutionary timescale will become longer. Most sources belong to classes (iii) and (iv), while only two direct impact systems and four persistent high state sources are currently known. The transition between groups (ii) and (iii) is thought to occur around P$_{\text{orb}}$\,$\sim$\,20 min, but the exact moment at which the transition occurs is currently not well constrained. 

\citet{Tsugawa1998} present a disc instability model for He-rich accretion discs in AM CVns (see also \citeauthor{Kotko2012} \citeyear{Kotko2012}), in analogy with the H-rich discs in cataclysmic variables \citep{Smak1982}. They showed that, depending on the accretion rate and accretor mass (which set the temperature at the outer and inner edges of the disc, respectively), three distinct situations can occur: if the mass transfer rate is very high, the material in the disc will be fully ionized and the disc is in a hot, stable state. If the mass transfer rate is very low, no ionized material is present in the disc and the system is in a cool, stable state. For mass transfer rates in between these two limiting cases, the temperature will increase as more material is transferred from the donor to the disc. When the temperature surpasses the ionization temperature of He, this leads to an increase in the viscosity in the disc. When this thermal instability occurs, the material in the disc is rapidly accreted onto the WD and an outburst is observed. Other system properties such as the chemical composition will influence whether the disc remains stable or goes into outburst \citep{Kotko2012}.

In this letter we present the identification of CXOGBS J175107.6-294037 (also known and from here-on referred to as CX361) as a persistent high state AM CVn binary.

\section{Observations}
\label{sec:data}
\subsection{X-ray and UV observations}
CX361 is an X-ray source detected in Chandra observations taken as part of the Galactic Bulge Survey (GBS; \citeauthor{Jonker2011} \citeyear{Jonker2011}, \citeyear{Jonker2014}). 
We search the Chandra archive around the position of CX361 and find 14 additional observations of the region, taken between 2006 and 2009, with exposure times ranging from 11 to 164~ks (Table \ref{tab:xraydata}).
\begin{table}
 \centering
  \caption{Summary of Chandra observations of CX361, obtained up to 2009. The observation marked with an asterisk was performed as part of the GBS. R$_{95}$ denotes the 95 per cent uncertainty of the X-ray position. The source counts were extracted from a circular region with a radius of 11\arcsec. }
  \begin{tabular}{cccccc}
ObsID & RA ($^{\circ}$)	&	Dec ($^{\circ}$)& R$_{95}$ ($\arcsec$) & T$_{\text{exp}}$ (ks)& Counts 	\\
\hline
 5934& 	267.7819& 	--29.6770	    	& 0.68& 40.5 & 	216	\\
 6362& 	267.7819& 	--29.6771    	& 0.70& 37.7	&180	\\
 6365& 	267.7819& 	--29.6771    	&1.06& 20.7	 &90	\\
 7168& 	267.7820& 	--29.6770    	&0.57& 13.8	 &24	\\
8753* & 267.782	&--29.677 &  6.43 & 2.0 & 7 \\
  9500& 	267.7819& 	--29.6771    	&0.59& 162.6&700	\\
 9501& 	267.7819& 	--29.6771    	&0.60& 131.0&539	\\
 9502& 	267.7820& 	--29.6771    	&0.59& 164.1&755	\\
 9503& 	267.7819& 	--29.6771    	&0.61& 102.3&468	\\
 9504& 	267.7819& 	--29.6770    	&0.61& 125.4&474	\\
 9505& 	267.7820& 	--29.6772    	&1.82& 10.7	 &39	\\
 9854& 	267.7820& 	--29.6770    	&0.98& 22.8	&112\\
 9855& 	267.7818& 	--29.6770    	&0.73& 55.9	&291	\\
 9892& 	267.7819& 	--29.6770    	&0.65& 65.8	&287	\\
 9893& 	267.7820& 	--29.6772    	&0.70& 42.2	&174	\\

  \end{tabular}
  \label{tab:xraydata}
\end{table}

The field of CX361 was observed in the near-UV (NUV) using the Galaxy Evolution Explorer (Galex; \citeauthor{Morrissey2007} \citeyear{Morrissey2007}) on 2011 July 1. The NUV filter covers the wavelength range between 1771 and 2831\,\AA, with an effective wavelength of 2316\,\AA. The exposure time is 586\,s, resulting in a 5$\sigma$ depth of $m_{NUV}$\,=\,20.3 mag. The full-width at half maximum of the point spread function is 5\farcs6.
\subsection{Optical observations}
\subsubsection{Photometry}
\citet{Wevers2016a} present optical imaging observations of CX361, taken as part of the GBS. The images were taken in three filters: $r^{\prime}$, $i^{\prime}$ and H$\alpha$. The field of CX361 was observed on 2006 June 29. The mean magnitudes are $r^{\prime}$\,=\,17.62\,$\pm$\,0.03, $i^{\prime}$\,=\,17.56\,$\pm$\,0.04 and H${\alpha}$\,=\,17.42\,$\pm$\,0.03.

The optical counterpart to CX361 was observed from 2010 June 12--18 with the MOSAIC--II instrument mounted on the Victor M. Blanco telescope to search for photometric variability \citep{Britt2014}. The field was covered on average 4 times per night using the SDSS $r^{\prime}$-band during 7 consecutive nights. 

The field of CX361 was also covered in the Optical Gravitational Lensing Experiment (OGLE) III (BLG194.1.659) and IV (BLG501.22.94718) surveys \citep{Udalski2015}, providing us with a lightcurve including 14298 observations in the \textit{I}-band, starting in 2001 and spanning 15 years. The cadence of these observations ranges from once every night up to every 20 minutes, and exposure times are 120 and 100~s for OGLE III and IV observations, respectively. The typical photometric uncertainty is 15 mmag.

\subsubsection{Spectroscopy}
We observed the optical counterpart to CX361 on 2012 May 25 using the Goodman high-resolution spectrograph mounted on the Southern Astrophysical Research (SOAR) telescope. We used the 400l/mm grating in combination with a 1\arcsec$\ $slit in 1\arcsec$\ $seeing, covering wavelengths between 3000 and 7000\,\AA. The exposure time was 1200~s. The spectrum has a dispersion of 1\,\AA/pix and the spectral resolution is R$\sim$1850. No flux standard was observed. The spectrum was reduced and extracted with standard tasks in \textsc{iraf}. 

We also present a spectrum taken with the VIsible MultiObject Spectrograph (VIMOS) mounted on VLT-UT3. Two 875~s exposures were obtained on 2013 July 6, using the medium-resolution MR grism in combination with the GG4750 order sorting filter, covering the wavelength range between 4800 and 10000\,\AA$\ $with a dispersion of 2.5\,\AA/pix. The seeing was 1.2\arcsec, and the 1\arcsec$\ $slit delivered a spectral resolution of R$\sim$600. The data were reduced, combined and extracted following the steps in \citet{Torres2014}. In addition, we corrected the resulting spectrum for the instrumental response using the standard star LTT9491.

\section{Analysis and results}	
\label{sec:results}
We use \textsc{wavdetect} from the \textsc{CIAO} software tools \citep{Fruscione2006} v4.7 at the position of CX361 to recover the X-ray source in the archival Chandra observations. We calculate the positional uncertainties following \citet{Evans2010}. We detect the source in all cases, with an average absorbed flux in the 0.3\,--\,7~keV range of 5.4$\,\pm\,$0.1$\times$10$^{-14}$~erg~cm$^{-2}$~s$^{-1}$. The best fit source positions are consistent within the errors, and the smallest X-ray error circle we obtain is 0\farcs57. Using \textsc{xspec} v12.9.0, the best combined fit to the spectra (comprising 2183 bins containing at least 1 photon) is a power law with an index of 2.10$\,\pm\,$0.06 and a hydrogen column density of n$_\text{H}$ = 1.5$\,\pm\,$0.2$\times$10$^{21}$~cm$^{-2}$. We use \textit{cstat} in \textsc{xspec}, which calculates C-statistics modified for the background \citep{Cash1979}, and find W\,=\,1809 for 2180 degrees of freedom.

\begin{figure} 
\centering
  \includegraphics[height=6cm, keepaspectratio]{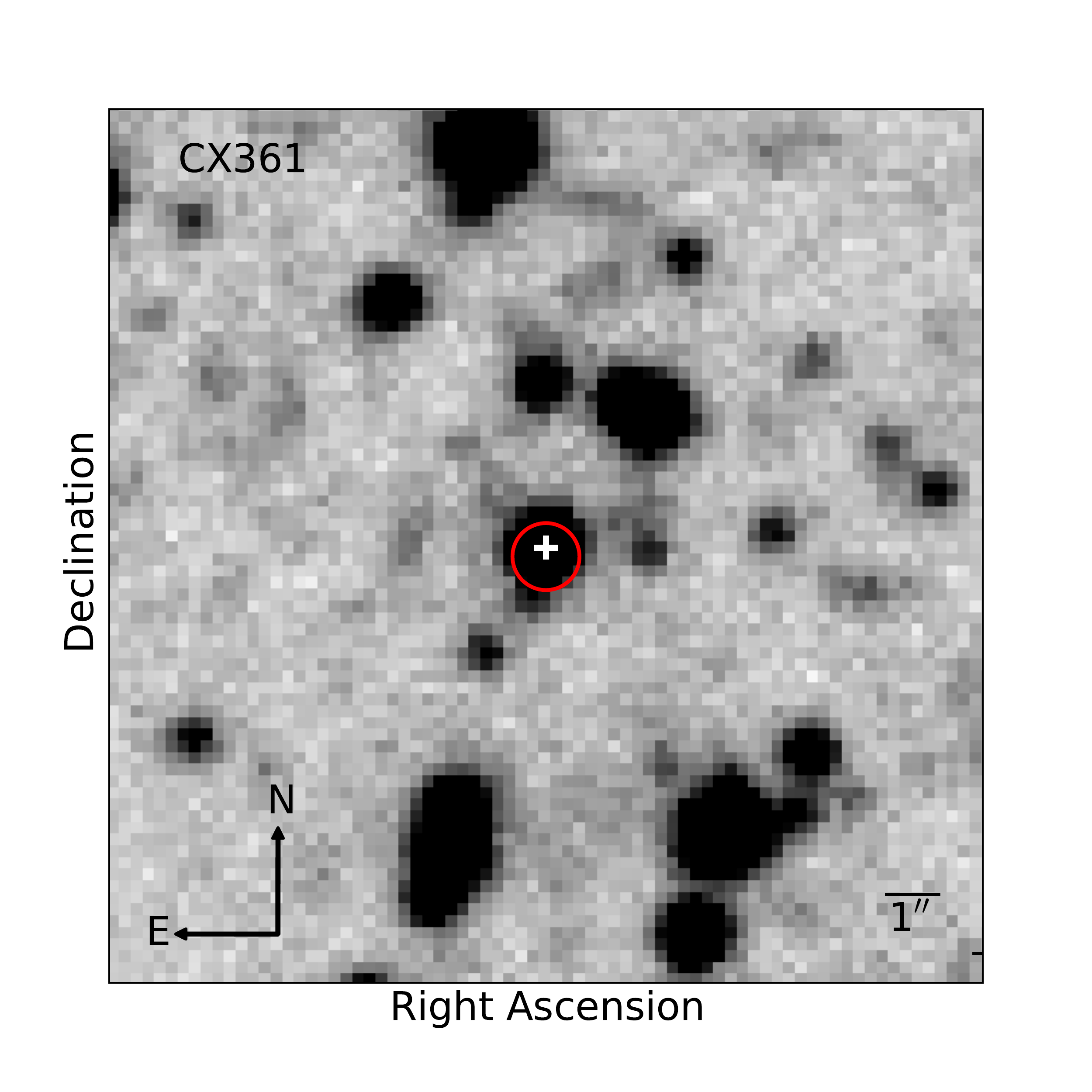}
  \caption{VLT/VIMOS image of CX361 in the $r^{\prime}$-band. The red circle indicates the best X-ray localisation, the white cross marks the optical counterpart.}
  \label{fig:finderchart}
\end{figure}

The small X-ray positional uncertainty allows us to identify the unique optical and UV counterpart. Figure \ref{fig:finderchart} shows an $r^{\prime}$-band image taken at the VLT. The system has a blue $r^{\prime}$-$i^{\prime}$ colour with respect to other stars in the field, and the NUV detection (m$_{NUV}=18.33$ mag) confirms that it has an intrinsically blue colour. We calculate an X-ray to optical flux ratio of log\Big($\frac{\text{F}_x}{\text{F}_{\text{opt}}}$\Big) = -1.26 using the $i^{\prime}$-band magnitude from \citet{Wevers2016a}.

\begin{figure*}
  \includegraphics[height=6.cm, keepaspectratio]{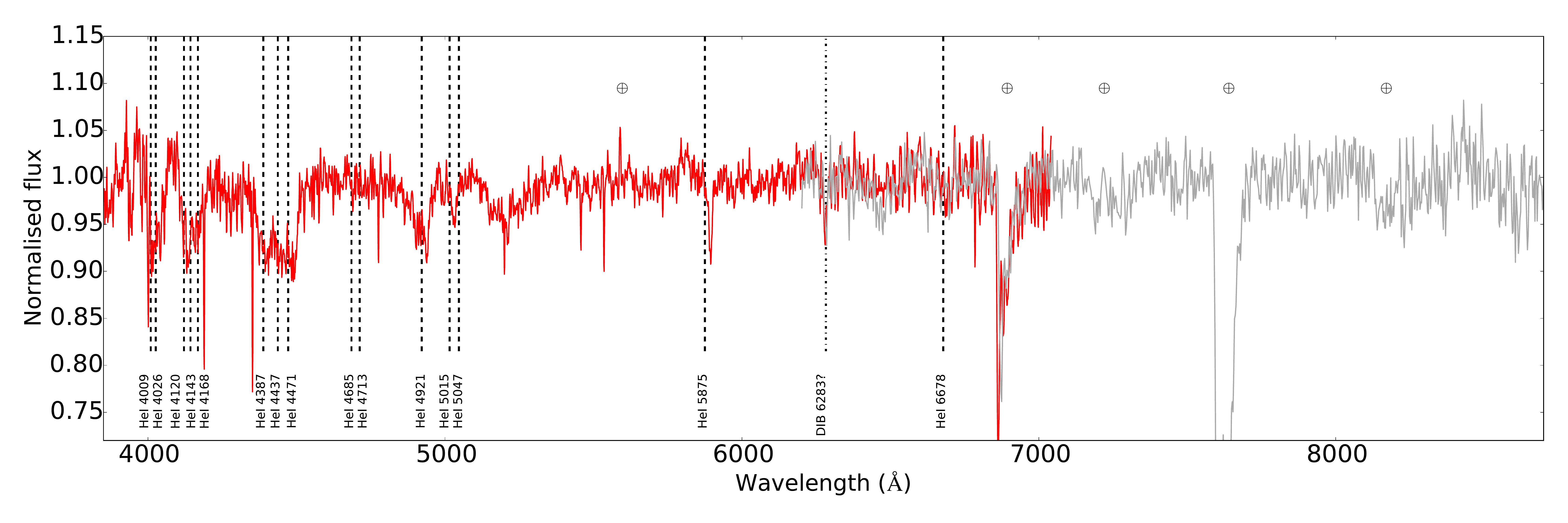}
  \caption{SOAR (red) and VIMOS (grey) spectra of CX361, normalised to the local continuum. Broad, shallow and asymmetric He\,\textsc{i} absorption lines are evident in the spectrum (marked by dashed lines). Because of the low signal-to-noise ratio, we apply a boxcar filter with a width of 3 pixels to smooth the SOAR spectrum for display purposes. Telluric features are marked by Earth symbols.}
  \label{fig:opticalspectra}
\end{figure*}

The two optical spectra of CX361, taken about a year apart, are presented in Figure \ref{fig:opticalspectra}. We normalise both spectra to the local continuum by fitting cubic splines in \textsc{molly}. The spectrum shows broad, shallow He\,\textsc{i} absorption lines (marked by dashed lines). No signs of hydrogen are present. The absorption feature at 5893\,\AA$\ $is likely due to a blend of He\,\textsc{i} 5875\,\AA$\ $with the Na\,\textsc{i} D doublet at 5890+5895\,\AA. The absorption feature at 6283\,\AA$\ $could be a diffuse interstellar band (DIB) or telluric in nature. We only show the red part of the VIMOS spectrum (grey) because it is contaminated by instrumental artifacts blueward of 6000\,\AA \footnote{http://www.eso.org/sci/facilities/paranal/instruments/vimos/doc/}. We search for absorption lines due to metals, but do not find any. 


The long term \textit{I}-band lightcurve of CX361 is shown in Figure \ref{fig:oglelc}. The epochs of the spectroscopic observations are marked by two vertical dashed lines. We search for periodicities by computing a Lomb-Scargle (LS) periodogram with periods ranging from 5 to 40 minutes with a 1s resolution. Because the cadence of the OGLE III data is low, we only use the detrended OGLE IV data for this analysis. We detrended the lightcurve by correcting for the difference in mean brightness between different seasons of OGLE IV. We find several peaks at periods ranging from 22 to 24 minutes (Figure \ref{fig:psd}). There are two sets of alias peaks visible in the periodogram, which we interpret as being two significant periodicities and their 1 day alias peaks. The periodicities with the most power are P$_1$\,=\,22.90\,$\pm$\,0.01 min and P$_2$\,=\,23.22\,$\pm$\,0.01 min (marked by red arrows). The period uncertainties are calculated by bootstrapping the lightcurve with resampling, and taking the standard deviation of the frequency distribution for each peak. We determine the (white) noise level of the LS amplitude $\sigma$ by randomising the observations in time and calculating the LS amplitude of the highest peak for 1000 trials. $\sigma$ represents the average of this amplitude distribution. An analysis performed on each season of OGLE IV observations separately confirms that the signals are always present. Although the Mosaic--II data consists of only 31 $r^{\prime}$-band datapoints spanning 7 nights, we recover two peaks at the same periods as found in the OGLE data: P\,=\,22.90 min and its 1 day alias at 23.28 min. 

\section{Discussion and conclusion}
\label{sec:discussion}
We have discovered a persistent X-ray source, CX361, in the Galactic Bulge Survey. We obtained optical spectroscopy, presented in Figure \ref{fig:opticalspectra}. The spectrum shows shallow, broad and asymmetric He\,\textsc{i} absorption lines, lacking signatures of hydrogen. This is typical for AM CVn binaries in a high accretion state. The spectrum looks very similar to high state systems such as AM CVn itself \citep{Roelofs2006} and HP Lib \citep{Roelofs2007}. 

\begin{figure}
  \includegraphics[height=4.4cm, keepaspectratio]{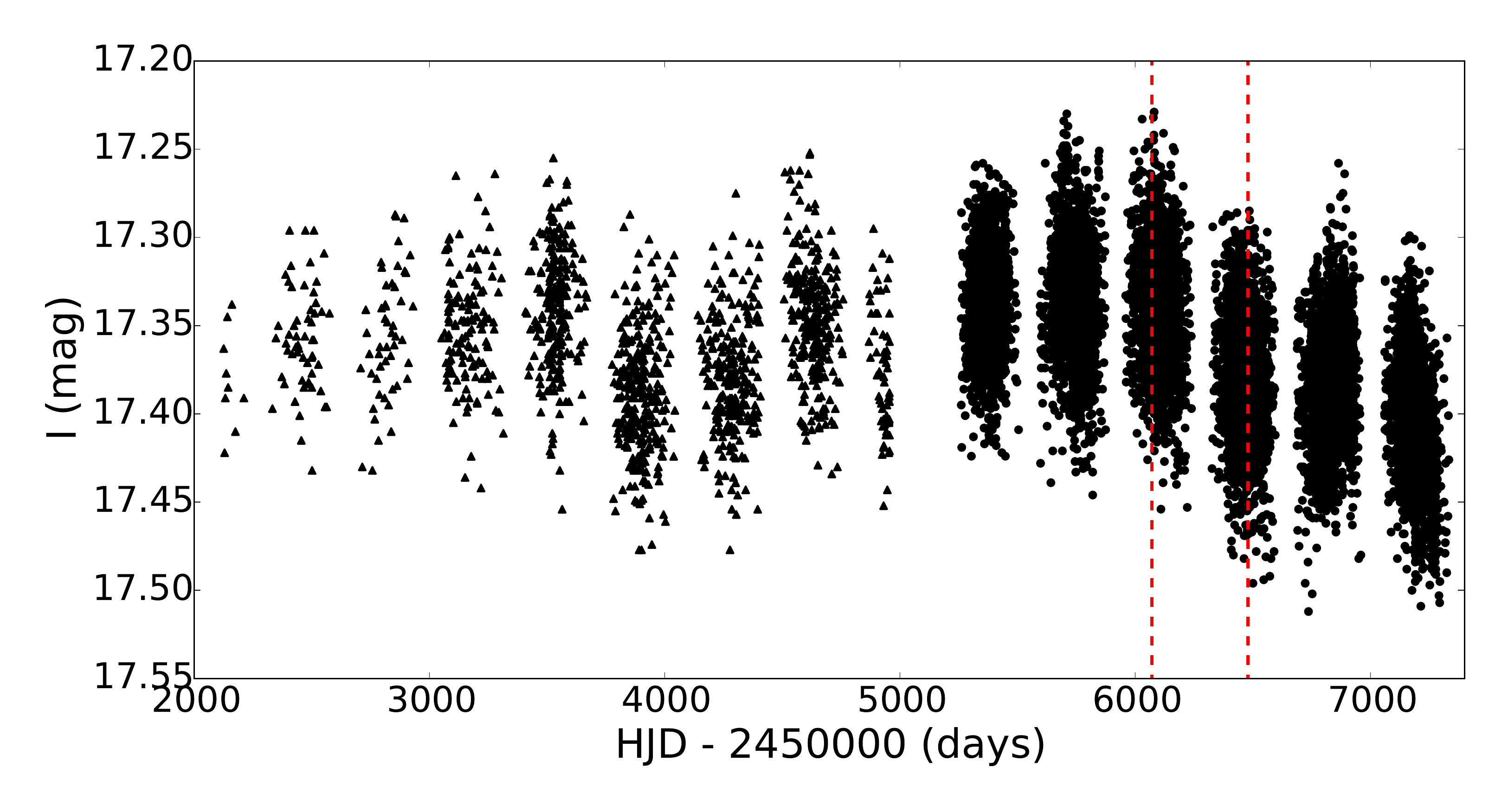}
  \caption{Long term \textit{I}-band lightcurve of CX361 from OGLE III (t\,$\leq$\,5000) and IV (t\,$\geq$\,5000). Variability is present at the 0.2 mag level on hours, days and weeks timescales. There is also evidence for a long term modulation of the brightness. The dashed lines mark the epochs of the SOAR (left) and VIMOS (right) spectra.}
  \label{fig:oglelc}
\end{figure}

Small amplitude variability is observed in the optical lightcurve on timescales of hours, days and weeks at a level $\sim$0.2 mag (Figure \ref{fig:oglelc}). This is consistent with an origin in an accretion disc, and similar to the photometric properties of high state AM CVn binaries \citep{Skillman1999, Patterson2002}. Our photometric observations also show that no outbursts have been detected in the last 15 years. Because the typical outbursts of an AM CVn system with P$_{\text{orb}} \sim$ 23 min (see below for an explanation) last 15 days, recur every 40 days and have an amplitude of $\sim$3 mag \citep{Levitan2015}, these would have been detected in our observations. Given that the spectrum indicates the system is in a high state, the lack of large amplitude variability suggests that the disc is persistently in the high state. On top of the flickering, there is evidence for a long term modulation of the brightness on a timescale of years. This could indicate deviations from the secular mass transfer rate set by the GWR, or due to variable irradiation of the donor by disc precession and/or warping \citep{Kotko2012}.

\begin{figure} 
  \includegraphics[height=8.5cm, width=0.5\textwidth]{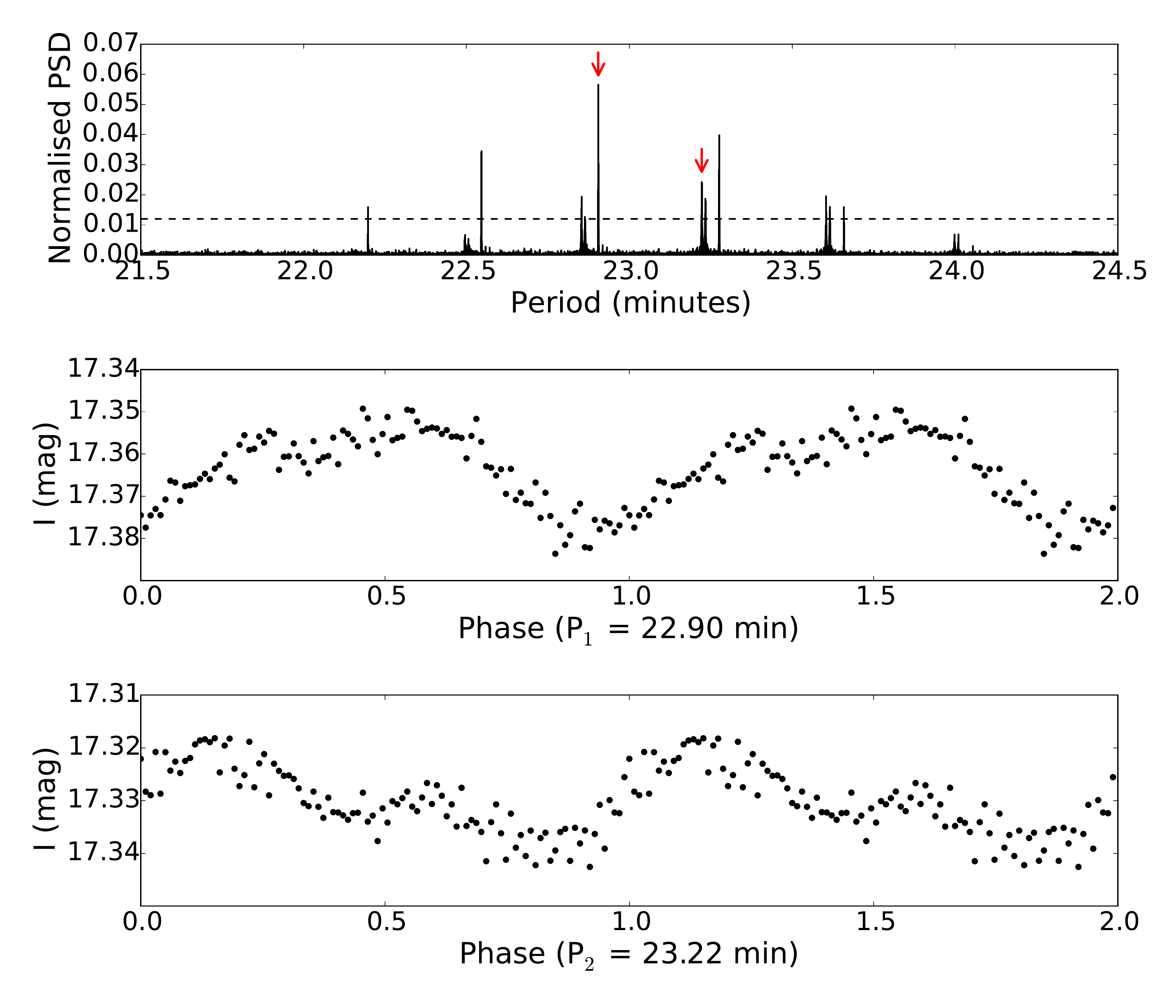}
  \caption{Top: LS periodogram of the detrended OGLE IV lightcurve. The arrows indicate two periodicities 22.90 and 23.22 min. The 5$\sigma$ detection level is marked by a dashed line. Middle: OGLE lightcurve, binned and folded on the 22.90 min period. Bottom: detrended OGLE IV lightcurve, binned and folded on the 23.22 min period. The zero phase is arbitrary.}
  \label{fig:psd}
\end{figure} 

A Lomb-Scargle analysis of the optical lightcurves shows evidence for two significant periodicities at P$_1$\,=\,22.90\,$\pm$\,0.01 and P$_2$\,=\,23.22\,$\pm$\,0.01 min (Figure \ref{fig:psd}). Their 1 day aliases are clearly visible in the periodogram. Typically, two strong periodicities are discovered in high state AM CVn sources: one at the orbital period and one at the superhump period. These periods are closely related, as the superhump period is likely caused by periodically modulated dissipation of the kinetic energy of the accretion stream due to variable irradiation \citep{Smak2009, Smak2013}. The photometric period may be related to the orbital period \citep{Levitan2011}, but this does not hold for all systems (see e.g. \citeauthor{Morales2003} \citeyear{Morales2003}). 

We show the binned lightcurve of the OGLE data folded on P$_1$ and P$_2$ in the middle and bottom panels of Figure \ref{fig:psd}, respectively. For P$_1$, the amplitude of variability over one cycle is $\sim$\,0.03 mag with a typical scatter of $\sim$\,0.01 mag. For P$_2$, the amplitude is 0.02 mag with 0.01 mag scatter. Comparing the morphology of these phase folded, binned lightcurves to those of HP Lib \citep{Patterson2002} and AM CVn \citep{Skillman1999}, we conclude that P$_1$ is likely the orbital period and P$_2$ the superhump period. We note that folding the entire OGLE lightcurve on P$_2$ does not result in a clean periodicity. For that reason, we show the detrended OGLE IV data folded on P$_2$ instead. This could suggest that there are slight variations in the superhump period over time. 

The difference between P$_1$ and P$_2$ is 19~s. Assuming that P$_1$\,=\,P$_{\text{orb}}$ and P$_2$\,=\,P$_{\text{sh}}$, this leads to a period excess ($\epsilon$\,=\,$\frac{P_{\text{sh}}\,-\,P_{\text{orb}}}{P_{\text{orb}}}$) of $\epsilon$\,=\,0.0138, similar to that of HP Lib ($\epsilon$\,=\,0.0148). This is consistent with the (tentative) relation between $\epsilon$ and P$_{\text{orb}}$ for systems with a dynamically determined orbital period (fig. 27 in \citeauthor{Solheim2010} \citeyear{Solheim2010}).

We can infer a mass transfer rate of $\dot{M}$\,=\,10$^{-9}$~M$_{\odot}$~yr$^{-1}$ by using the relation between P$_{\text{orb}}$ and $\dot{M}$ found by \citet{Cannizzo2015}. Comparing this with the region of He disc stability (\citeauthor{Tsugawa1998} \citeyear{Tsugawa1998}; see also fig. 2 in \citeauthor{Deloye2005} \citeyear{Deloye2005}), we see that the system falls in the stable disc regime, consistent with our interpretation. 

We rule out a scenario with an ultra-compact X-ray binary (UCXB) in outburst based on the low observed $\frac{\text{F}_x}{\text{F}_{\text{opt}}}$ ratio. \citet{Jonker2011} calculate the typical X-ray fluxes and optical magnitudes for a population of AM CVns and UCXBs based on observed properties. Comparing our measurements to their populations (fig. 4 in \citeauthor{Jonker2011} \citeyear{Jonker2011}), we see that they are inconsistent with an UCXB scenario. In that case, the system should have at least an order of magnitude higher $\frac{\text{F}_x}{\text{F}_{\text{opt}}}$ ratio than observed.

Phase-resolved spectroscopy is required to unambiguously determine the orbital period. If confirmed at 22.90 min, the orbital period would be similar to that of a confirmed outbursting system (22.5 min for PTF1 J191905.19+481506.2; \citeauthor{Levitan2014} \citeyear{Levitan2014}). This would provide direct observational evidence that in addition to P$_{\text{orb}}$, other factors play an important role in the disc stability of high state AM CVn systems. Comparing our values of $\dot{M}$ and P$_{\text{orb}}$ to fig. 2 of \citet{Deloye2007conf}, we conclude that our estimates are inconsistent with a zero-temperature WD but rather suggest a semi-degenerate donor star (either a high-entropy WD or He star). The implied high-entropy donor star may in part explain why the disc in this system is stable while PTF1 J1919, at similar P$_{\text{orb}}$, shows outbursts. 

Using the X-ray observations, we can obtain two distance estimates for the system. By using the relation between the hydrogen column density (N$_H$, determined from our fits to the X-ray spectra) and optical extinction found by \citet{Guver2009}, we estimate that the extinction in the \textit{V}-band is A$_V$\,=\,0.7 mag. Using the 3D reddening map by \citet{Green2015}, this corresponds to a distance of $\sim$ 500 pc. Alternatively, assuming that the X-ray luminosity is equal to that of HP Lib (1.4\,$\times$\,10$^{31}$~erg~s$^{-1}$; \citeauthor{Ramsay2006} \citeyear{Ramsay2006}) we estimate a distance to CX361 of 1.1 kpc. 

A short orbital period would make this source an excellent candidate to serve as a verification source for the future \textit{eLISA} space-based gravitational wave mission \citep{Amaro2012}. Assuming AM CVn-like parameters, but scaling for the tentative orbital period and distance of CX361 we can get a rough estimate of the GW strain (\textit{h}). Using \textit{h}\,=\,2.1$\times$10$^{-22}$ for AM CVn \citep{Roelofs2006}, we find \textit{h}\,$\sim$\,9\,$\times$\,10$^{-23}$ at 1.1 kpc, roughly a factor 2 above the confusion-limited Galactic background predicted by \citet{Nelemans2004}. A determination of system parameters such as an accurate distance and inclination are needed to confirm the expected GW strain caused by this source.

\section*{Acknowledgements}
We thank Tom Marsh for developing the software package \textsc{molly}. 
PGJ and ZKR acknowledge support from European Research Council Consolidator Grant 647208. COH acknowledges support from an NSERC Discovery Grant, and Discovery Accelerator Supplement. The OGLE project has received funding from the National Science Centre, Poland, grant MAESTRO 2014/14/A/ST9/00121 to AU. J.W. and R.I.H. acknowledge support from the National Aeronautics and Space Administration through Chandra Award Number AR5-16004X issued by the Chandra Xray Observatory Center, which is operated by the Smithsonian Astrophysical Observatory for and on behalf of the National Aeronautics Space Administration under contract NAS8-03060. Based in part on data obtained from the Chandra and Galex Data Archives.


\bibliographystyle{mnras.bst}
\bibliography{bibliography_cx361.bib}

\label{lastpage}

\end{document}